\documentclass[showpacs, twocolumn,preprintnumbers,secnumarabic,amsmath,amssymb,10pt, nofootinbib]{revtex4-1}
\usepackage{dcolumn}     
\usepackage{bm}   
\pdfoutput=1        
\usepackage{graphicx}
\newcommand{\be}{\begin{equation}}

\newcommand{\ee}{\end{equation}}
\newcommand{\bea}{\begin{eqnarray}}
\newcommand{\eea}{\end{eqnarray}}

\begin{document}
\title{The Thermodynamics of Relativistic Multi-Fluid Systems}
\author{Bob Osano$^{1,2}$ \\
\small{$^{1}$Centre for Higher Education Development, \\University of Cape Town (UCT), Rondebosch 7701,Cape Town, South Africa}\\$\&$\\
\small{$^{2}$Cosmology and Gravity Group, Department of Mathematics and Applied Mathematics,\\ University of Cape Town (UCT), Rondebosch 7701, Cape Town, South Africa\\}}

\begin{abstract}
This article extends the single-fluid relativistic irreversible thermodynamics theory of {\it M{\"u}ller}, {\it Israel} and {\it Stewart} (hereafter the {MIS} theory) to a multi-fluid system with inherent species interactions. This is illustrated in a two-fluid toy-model where an effective complex 4-velocity plays the role of a primary dynamical parameter. We find that an observer who resides in the {\it real}-part of this universe will notice that their knowledge of the universe parametrized using {\it real}, rather than {\it imaginary}, quantities are insufficient to fully determine properties such as the total energy density, pressure or entropy,  In fact, such an observer will deduce the existence of some negative energy that affects the expansion of their perceived {\it real} universe.
\end{abstract}
\nopagebreak
\date{\today}
\maketitle
\section{Introduction}
The extended relativistic thermodynamics theory by {\it M{\"u}ller}, {\it Israel} and {\it Stewart} developed in \cite{Isra0, Isra1, Muller1, Rug} has found wide application in scenarios where the material content under investigation that can be modelled using single-fluid approximation, whether the material is made of (1) one species or (2) several species whose properties are given by the average or the bulk behaviour. This treatment forms the foundation of most studies in relativistic non-equilibrium thermodynamics found in the literature for idealised fluids. How closely this approximation models realistic fluids mixture needs to be investigated. To this end, we formulate a multi-fluid approximation as the first step in the relaxation of the conditions governing the $MIS$ theory. We will demand $MIS$ theory is recovered in the appropriately motivated limit. Part of the requirement will be that the resulting system of differential equations must retain their hyperbolicity \cite{Geroch1, Rug} in line with the principle of causality and be stable \cite{Prou1} for the formulation to be viable and predictive \cite{BobTim}. 

The aim of this article is modest and is primarily to present a formalism for the thermodynamics of relativistic system of multi-fluid. Since we intend to extend the $MIS$ theory, we will begin with the assumption that the $MIS$ model is the standard theory and extension must necessarily recover it when subjected to physically motivated constraints or conditions that impact the nature of fluid approximation. Several properties that are established in the single-fluid approximation suddenly lose clarity. In terms of thermodynamics, these are the definitions of  (i) a universal temperature\cite{Lands1, Lands2}, (ii) entropy, (iii) heat, and  (iv) work, to mention but the primary one. This lack of clarity impacts the laws of thermodynamics and will require scrutiny.

The problem of defining a universal {\it temperature} was first encountered in the none-equilibrium thermodynamics theories in the single-fluid approximation. At the core of the problem is the seeming non-existence of a Lorentzian type transformation between reference frames that readily recovers the black-body temperature given the necessary constraints \cite{Lands1}. Some progress has been made in this regard, see \cite{Lands2} for example, but the debate is not closed. It is important to emphasise that we will make some assumption in this paper regarding how a temperature transforms, to allow for progress but the description should not be taken as definitive. A detailed analysis of such transformations will be pursued in future \cite{ Bob3}. As for the definition of {\it entropy}, it is known that when solving gravitational field equations, the standard approach takes into account the bulk-effects and ignores the surface effects. But it is known that when surface terms are evaluated at the horizon they give the entropy of such a horizon ( see \cite{Pad3} and references therein). This suggests that entropy should ideally include the body and surface terms.

The characterization of $heat$ and $work$ are also less straight forward, see \cite{Nak}. It is known that {\it the heating, however, heat is defined, of a space-time that is endued with a certain microscopic degree of freedom and which is capable of exhibiting thermal phenomena will necessarily create micro-structure}. It is sensible to consider the converse of this and therefore ask how do micro-structures affects the macro-properties in space-time dynamics? It is this notion of micro-structure that motivates the formulation that we develop in this article.

This article is organised as follows: section (\ref{sec1}) discusses the thermodynamics for single-fluid approximation while section(\ref{sec2}) discusses thermodynamics for multi-fluid approximation. Section (\ref{sec3}) discusses the special case of two-fluid approximation. Section (\ref{sec4}) gives the discussion and the conclusion.

\section{Irreversible Thermodynamics and the Single-fluid approximation}\label{sec1}
In order to understand the complicated nature of multi-fluid dynamics, it is useful to review the single-fluid approximation in the context of general relativity and cosmology. 
Early theories on single-fluid irreversible thermodynamics \cite{Eckart, Klutten1, Klutten2, Landau, Cattaneo, Kranyz, Aitken, Kampen, Stewart3} were plagued by the pathology that they predicted instantaneous propagation of viscous and thermal effects, given the parabolic nature of the resultant differential equations. This made such theories to be predictive only for slowly varying systems. The pathology was traced, see \cite{Isra0}, to the {\it non-perturbative} \cite{Bob3} truncation procedure which led to the dropping of quadratic terms from the heat and viscous stresses in the expression for the entropy 4-vector. The entropy, in this case, includes both the material and 4-momentum fluxes. This was clearly not suitable for fast varying systems and a new theory was therefore required. This led to the development of the theory that we will discuss next.

\subsection{The M\"{u}ller-Israel-Stewart ($MIS$) theory }Also called {\it Israel-Stewart} theory, the $MIS$ theory is a theory for relativistic irreversible thermodynamics based on single-fluid approximation. The state of fluid is generally given by three entities; the stress-energy-momentum tensor, $T^{\mu\nu}$, the particle flux $N^{\mu}$ and the entropy flux $S^{\mu}$. The momentum tensor and the particle flux obey their respective conservation laws, ${T^{\mu\nu}}_{;\nu}=0$ and ${N^{\mu}}_{;\mu}=0,$ while the entropy vector obeys the second law of thermodynamics ${S^{\mu}}_{;\mu}> 0$. The semicolon denotes the covariant derivative. If $u^{a}$ is a time-like vector and $h_{\mu\beta}$ is an orthogonal projection tensor, it is follows that the energy density $\rho=T^{\mu\beta}u_{\mu}u_{\nu}>0.$  $T^{\mu\beta}$ has a time-like unit vector $u^{\mu}_{\vert_E}$ (i.e.  $ u^{\mu}_{\vert_E} u_{\mu_{\vert_{ E}}} =-1$). This is not the only time-like unit vector that one can find for such a fluid. In fact, one can define other unit vectors for example, $u^{\mu}_{\vert_{N}}=N^{\mu}/{\sqrt{-N^{\mu}N_{\mu}}}$ or even $u^{\mu}_{\vert_{S}}=S^{\mu}/{\sqrt{-S^{\mu}S_{\mu}}}$. It follows that $u^{\mu}_{\vert_{N}}=u^{\mu}_{\vert_{E}}$ for fluid at equilibrium, which suggests that there exists a unique time like 4-velocity vector which will denote by $u^{\mu}.$ The full set of relevant equations for the perfect fluid case, synonymous with the thermodynamic equilibrium, take:
\bea
T^{\mu\nu}&=& \rho u^{\mu}u^{\nu}+ph^{\mu\nu}\\
N^{\mu}&=&nu^{\mu}\\
S^{\mu}&=&su^{\mu},
\eea where $p=T^{\mu\nu}h_{\mu\nu}$ is the isotropic pressure, $\rho$ is again the energy density, $n$ is the number density, $s$ is the flux density. An alternative formulation, in terms of divergence type functions, is given in the appendix (\ref{div}). In the present formulation, the equation of state is given by $p=p(\rho,s)$.

Deviation from equilibrium can then be characterised using \bea \label{drfteq} u^{\mu}_{\vert_{N}}-u^{\mu}_{\vert_{E}}=\mathcal{V}^{\mu}\neq 0\eea where the requirement that $\mathcal{V}^{\mu}<<1$ may be used to characterise {\it close to}-or{\it quasi}-equilibrium \cite{Salaz}. In this case one can define an orthogonal projection tensor $h_{\mu\beta}=g_{\mu\beta}+u_{\mu}u_{\beta}$.This is because of \bea
{u^{\mu}}_{\vert_{N}}{u_{\mu\vert_{N}}}&=&{\mathcal{V}^{\mu}}{\mathcal{V}_{\mu}}+\mathcal{V}^{\mu}{u_{\mu\vert_{N}}}+{u^{\mu}}_{\vert_{N}}{\mathcal{V}_{\mu}}+{u^{\mu}}_{\vert_{E}}{u_{\mu\vert_{E}}}\nonumber\\&=&-1,
\eea when evaluated in the reference frame of a perfect fluid. Since $$u^{\mu}_{\vert_{N}}=u^{\mu}_{\vert_{E}}+\mathcal{V}^{\mu},$$ $\mathcal{V}^{\mu}$ can be thought of a linear perturbation. ${\mathcal{V}^{\mu}}{\mathcal{V}_{\mu}}$ is a product first-order perturbation terms and therefore a higher order perturbation. Similarly, $\mathcal{V}^{\mu}{u_{\mu\vert_{N}}}$ is first order. These terms are dropped when evaluating in the frame. As in the case or perfect fluids, we do not have a unique 4-velocity vector for any arbitrary non-perfect fluid. A thermodynamics formulation for such non-prefect fluid must necessarily incorporate anisotropic stress and where appropriate heat exchange. These are characterised using the quantities $\pi_{\mu\nu}$, the anisotropic stress tensor, and $q_{\mu}$ , the energy flux vector respectively.

An appropriate starting point in the formulation of a theory of irreversible thermodynamics is to allow key parameters such as entropy and energy-momentum tensor to be functions of a broader number of properties, over above the standard volume and internal energy. To formalise such extension, let these properties be given by the generic scalar $f$, vector $f^{\mu}$ and tensor $f^{\mu\nu}$, in which case $s=s(f, f^{\mu}, f^{\mu\nu})$. These tensors of ranks 0,1 and 2, can be explicitly defined to have the physical meaning as discussed in \cite{Salaz, BobTim, BobTemp}. We note here that the tensors denote both bulk and surface terms. This is important as it will allow our description to include surface entropy.
It follows that \bea\label{eqent}
ds=\frac{\partial s}{\partial f}d f+\frac{\partial s}{\partial f^{\mu}}d f^{\mu}+\frac{\partial s}{\partial f^{\mu\nu}}d f^{\mu\nu},
\eea
the depended on a rank 2 tensor is an extension and leads to a generalized Gibbs relation. We note that we could have more than one of these intrinsic properties characterized by a scalar, a vector or a tensor. For example, the internal energy and volume are both scalars while heat is a vector. The coefficients in Eq.(\ref{eqent}) can then be treated in the usual manner with the case of $\partial s/\partial E=1/T$, where $T=T(f, f^{\mu}, f^{\mu\nu})$ is a non-linear temperature. It suffices to say that it is easy to recover the standard Gibbs relation by restricting Eq. (\ref{eqent}) to scalars. For the standard set of properties, it can be shown that
\bea
S^{\mu}=su^{\mu}+\frac{1}{T}q^{\mu}-\mathcal{Q}^{\mu},
\eea
where $\mathcal{Q}$ denotes a collection of second-order terms and takes the form

\bea
\mathcal{Q}^{\mu}&=&\frac{u^{\mu}}{2T}\left[\beta_{0}\Pi^{2}+\beta_{1}q_{\mu}q^{\mu}+\beta_{2}\pi^{\mu\nu}\pi_{\mu\nu}\right]\nonumber\\&&~~~~~~~~~~~~~~~~~~
-\frac{1}{T}(\alpha_{0}\Pi q^{\mu}-\alpha_{1}\pi^{\mu\nu} q_{\nu}+\mathcal{F}),
\eea
where $\mathcal{F}$ is a function of energy density, isotropic pressure, energy flux and the symmetric shear tensor. For a detailed discussion of this term, the coefficients, $\beta_{0},\beta_{1},\beta_{2},\alpha_{0},\alpha_{1}$ and those embedded in $\mathcal{F}$, the reader is referred to \cite{Isra0,Isra1} and for a pedagogical presentation to \cite{Salaz}. It follows that stress-energy momentum tensor for such non-perfect fluid takes the form
\bea
T^{\mu\nu}&=& \rho u^{\mu}u^{\nu}+ph^{\mu\nu}+2q^{(\mu}u^{\nu)}+\pi^{\mu\nu},
\eea
and is the {\it M\"{u}ller-Israel-Stewart} theory \cite{Muller1, Isra0, Isra1}. The particles flux also has a contribution from possible particle drift, $N^{\mu}=nu^{\mu}+n^{\mu}$, $n^{\mu}$ is the particle drift in the frame defined by $u^{\mu}.$ Two special frames in which the fluid dynamics has physical meaning, but not the only ones, are the $Landau-Liftshitz$ (energy) frame defined by $u^{\mu}=u^{\mu}_{\vert_{E}}\Rightarrow q^{\mu}=0$ and the $Eckart$ (particle) frame $u^{\mu}=u^{\mu}_{\vert_{N}}\Rightarrow n^{\mu}=0$. These are related via equation (\ref{drfteq}). The existence of different frames, in single-fluid approximation, implies the existence of different projection tensors whose generic form is $h_{\mu\nu}=g_{\mu\nu}+u_{\mu}u_{\nu}$. The frames of reference increase when one moves from single- fluid and multi-fluid approximation. Each constituent fluid will have a unique energy frame and a particle frame. We would now like to broaden this discussion by considering multiple fluids, where the above generalizations do not readily apply. We expect that there exist some limiting conditions that should allow us to recover the standard $MIS$ theory. The reader will remember that our standard concepts of $heat$ and $work$ are often frame depended. This means that the chose of a frame will determine our notion of these two concepts. It has been demonstrated \cite{Nak} that even the concept of volume may be frame depended leading to disparities in the estimation of fundamental quantities. This lack of clarity becomes even more pronounced when one considers more than one fluid but as we will show, one can develop global parameters that are linked to the local frame and which allow for ease of physical interpretation. This is pursued in the next section.
 
\section{ Irreversible Thermodynamics and the multi-fluid approximation}\label{sec2}  To make the approximation procedure that we will develop more explicit, it is helpful to borrow the language of fluid dynamics. In this regard, we need to discuss the scales in which fundamental changes take place when fluids mix and how these relate to the modelling techniques used. We adopt the language in \cite{Nguyen1}. It is known that when different fluids come into contact during their flows, the resultant mixture is in-homogeneous. However, the dynamical act of mixing is a transport process involving temperature, species, and phases that lead to a reduction of in-homogeneity. In an ordinary fluid, the mixing may generate other effects such as reaction or even changes in fluid properties. Mixing is often categorised as: {\it macro}, {\it meso} and {\it micro}. Whereas macro-mixing is governed by the largest scale of fluids motion, micro-mixing is governed by the opposite end of the scale; the smallest scale of fluids motion and even molecular motion. {\it In conventional macro-scale mixing, the smallest scale of fluid motion is the size of turbulent eddies, also called the Kolmogorov scale.} A scale of mixing that lies between macro and micro is referred to as {\it meso}-mixing. A theory of extended irreversible thermodynamics for single-fluid approximation based on the meso-scale was presented in \cite{Jou}. Taking some of these properties as analogues of cosmological fluids, we may want to identify a scale reminiscent to the Kolmogorov scale, which we will refer to as the cosmological Kolmogorov scale which would be a cut-off for approximating relativistic fluid properties. In this regard, we will not attempt to investigate relativistic turbulence instead but instead develop a formalism for two interacting fluids. Assume that the two fluids are denoted by $X$ and $Y$.The dynamical variables are the stress-energy tensors denoted by ($T^{\mu\nu}_{\vert_{X}},T^{\mu\nu}_{\vert_{Y}}$), the particle fluxes denoted by ($N^{\mu}_{\vert_{X}}, N^{\mu}_{\vert_{Y}}$) and entropy fluxes denoted by $S^{\mu}_{\vert_{X}}, S^{\mu}_{\vert_{Y}}$. To understand the conservation properties in a multi-fluid environment, it is important to distinguish between species interactions with dissipative properties (i) without entrainment, and (ii) with entrainment. In the first case  individual species obey own conservation laws and cumulative conservation i.e. if $f_{X}^{\mu\nu}$ and $f_{Y}^{\mu\nu}$ are tensor properties for the two species then $\nabla_{\mu}(f_{X}^{\mu\nu}+f_{Y}^{\mu\nu})=0$ which may be taken to imply $\nabla_{\mu}f_{X}^{\mu\nu}=0=\nabla_{\mu}f_{Y}^{\mu\nu}$. The case involving entrainment is more nuanced and $\nabla_{\mu}(f_{X}^{\mu\nu}+f_{Y}^{\mu\nu})\ne0$. In fact, $\nabla_{\mu}(f_{X}^{\mu\nu}+f_{Y}^{\mu\nu}+f_{XY}^{\mu\nu})=0$, where the last terms encode reaction and other non-standard interaction properties. We must consider the interaction components ($N^{\mu}_{\vert_{XY}}$, $T^{\mu\nu}_{\vert_{XY}}$and $S^{\mu}_{\vert_{XY}}$) in the multi-fluid formulation. Hence,  
\bea
\nabla_{\mu} \sum_{i}  N^{\mu\nu}_{\vert_{i}}&=&0=\nabla_{\mu} \sum_{i}  T^{\mu\nu}_{\vert_{i}},~~
\nabla_{\mu} \sum_{i}  S^{\mu}_{\vert_{i}}>0,
\eea
 where $i=X, Y $ or $XY$. The $XY$ incorporates {\it entrainment}, where the interaction allows for it \cite{Andersson1,BobTim}. It is instructive to note that two observers move with the 4-velocities $u^{\mu}_{\vert_{X}}(= N^{\mu}_{\vert_{X}}/ \sqrt{-{N^{\mu}}_{\vert_{X}}{N}_{\mu\vert_{X}}})$ and $u^{\mu}_{\vert_{Y}}(= N^{\mu}_{\vert_{Y}}/ \sqrt{-{N^{\mu}}_{\vert_{Y}}{N}_{\mu\vert_{Y}}})$ will have different rest-frames and different projections on their respective frames. These may be denoted by $h^{\mu\nu}_{\vert_{X}}=g^{\mu\nu}_{\vert_{X}}+u^{\mu}_{\vert_{X}}u^{\nu}_{\vert_{X}}$ and $h^{\mu\nu}_{\vert_{Y}}=g^{\mu\nu}_{\vert_{Y}}+u^{\mu}_{\vert_{Y}}u^{\nu}_{\vert_{Y}}$, with the special case $g^{\mu\nu}_{\vert_{X}}\equiv g^{\mu\nu}_{\vert_{Y}}$ (This is reminiscent of the $energy$ and the $particle$ frames occupying space with the same geometry). Although we have presented these as projections onto hyper-surfaces, they need not be so. Projection tensors could be surface-forming, such as the familiar case in general relativity which allows the curvature to be decomposed into equations that include the {\it Gauss-Codazzi equations}, or the {\it Gauss-Weingarten} relations linking embedded geometry connections to the hyper-surface geometry connections. None surface-forming projections also exist and allow for the definition of fluid properties such as shear and vorticity, and the familiar Raychaudhuri equations \cite{Gowdy1, Henk}. An alternative way to look at this is to consider one of the fluids, for example, that with the 4-velocity $u^{\mu}_{\vert_{X}}$. The condition $u^{\mu}_{\vert_{X}}{u_{\mu}}_{\vert_{X}}=-1$ suggests that existence of the projection tensor $U_{ab\vert{X}}= - {u_{\mu}}_{\vert_{X}}{u_{\nu}}_{\vert_{X}}$ that obeys the condition ${U^{\mu}}_{\gamma\vert{X}}{U^{\gamma}}_{\nu\vert{X}}={U^{\mu}}_{\nu\vert{X}}$. ${U^{\mu}}_{\nu\vert{X}}$ projects onto the tangent space of this fluid world-line. We demand that energy and the particle frame of the unified approach satisfy $|\hat{u}^{\mu}-\hat{u}^{\mu}_{E}|<<1$. Let us now define a resultant 4-velocity $\hat{u}^{\mu}=f(u^{\mu}_{\vert_{X}},u^{\mu}_{\vert_{Y}})$ and the corresponding projection tensor \bea
{{\hat{h}}^{\mu}}_{~\nu}={\hat{g}^{\mu}}_{~\nu}+\hat{u}^{\mu}\hat{u}_{\nu},
\eea that projects onto the rest-frame of the fluid mixture such that ${{\hat{h}}^{\nu}}_{~\mu}{{\hat{u}}}_{\nu}=0.$ We will assume that this observer, $\hat{u}^{\mu}$, is not accelerated in contrast to that considered in \cite{BobTim}. These velocity fields are chosen in such a way that they satisfy the concavity requirement \cite{Muller1}. Once the $\hat{u}^{\mu}$ is chosen, the observer moving with this velocity will record the energy density $\hat{\rho}$ and the particle flux $\hat{N}^{\alpha}=f({N^{\mu}}_{\vert{X}},{N^{\mu}}_{\vert{Y}})$ where $f$ denotes $'function~of'$.

 It follows that the total stress-energy momentum tensor is given by 
\bea
T^{\mu\nu}=\sum_{i}{T^{\mu\nu}}_{\vert{i}},\label{Emt1}
\eea where again $i=X,Y, XY$. These can be decomposed into the energy density, pressure ( isotropic and anisotropic) terms and the heat term in standard way and takes form:
\bea
T^{\mu\nu}_{\vert{i}}&=&\hat{\rho}_{\vert{i}}\hat{u}^{\mu}_{\vert{i}}\hat{u}^{\nu}_{\vert{i}}+\hat{p}_{\vert{i}}\hat{h}^{\mu,\nu}_{\vert{i}}+2\hat{u}^{(\mu}_{\vert{i}}\hat{q}^{\nu)}_{\vert{i}}+\hat{\pi}^{\mu\nu}_{\vert{i}},
\eea where the heat flux vector and is given by~$
\hat{q}_{\mu\vert{i}}=-\hat{h}_{\mu\nu\vert{i}}\hat{u}_{\gamma\vert{i}}T^{\nu\gamma}_{\vert{i}} $ while the anisotropic stress-energy tensor is given by $\hat{\pi}_{\mu\nu{\vert{i}}}=T^{\gamma\delta}_{\vert{i}}(\hat{h}_{\gamma\langle \mu}\hat{h}_{\nu\rangle \delta})_{\vert{i}}$ . The total entropy also takes the form
\bea
S^{\mu}=\sum_{i}{S^{\mu}}_{\vert{i}}+S_{surf},
\eea where again the individual contribution can be expressed in terms of heat vector, temperature and rest-frame defined entropy as will be shown in Eq. (\ref{en1}). $S_{surf}$ represents the entropy enveloping the volume occupied by the two fluids. The generic nature of the formalism presented in this section conceals its significance. We will remedy this situation, in the next section, by providing a fully worked out example. Let's look at it.
\section{The fundamental problem of two-fluid approximation: An illustration}\label{sec3} 
\begin{figure}[ht]
\begin{center}
\includegraphics[width=0.35\textwidth]{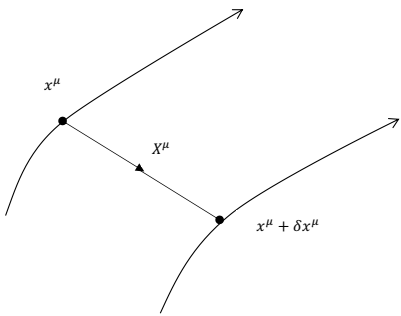} 
\caption{Co-moving observers and a connecting vector.\label{fig1}}
\end{center}
\end{figure}

Let us begin this section by briefly outlining what is meant by {\it neighbouring} word-lines, in the context of single-fluid approximation in cosmology. This preamble is necessitated by the need for clarity in discussing the differences between the {\it single-fluid approximation} and the {\it multi-fluid approximation} treatment that will later develop. Let $X^{\mu}$ be a vector whose components are given in a co-moving coordinate by
$
X^{\mu}=\delta x^{\mu}( X^{0}=0)$, which at all times joins the two world-lines given by $x^{\mu}$ and $x^{\mu}+\delta x^{\mu}$. The reader is referred to FIG. (\ref{fig1}) for a schematic representation of this set-up.
Since this is a co-moving system, there is one fundamental 4-velocity vector. The connector $X^{\mu}=(\partial x^{\mu}/\partial x^{\nu})\delta x^{\nu}$ does not lie in the rest-frame defined the 4-velocity. However, it is possible to define the time derivative and a relative position vector using the velocity vector and a projection tensor that projects relative to the rest-frame of such a velocity. i.e.
\bea
\dot{X}^{\mu}&=&u^{\mu}_{~;\nu}X^{\nu}\\ X^{\mu}_{\perp}&=&{h^{\mu}}_{\nu}X^{\nu}. \eea
 It follows that there exists a corresponding relative velocity vector $V^{\alpha}={V^{\alpha}}_{\beta}X^{\beta}_{\perp},$ where ${V^{\alpha}}_{\beta}={h^{\alpha}}_{\gamma}{h^{\delta}}_{\beta}u^{\gamma}_{~;\delta}$. This indicates that the relative velocity vector of the neighbouring co-moving  particles is linked to the relative position vector through a linear transformation as given by the spatial gradient of the 4-velocity vector. The relative vector can then be covariantly split into an expansion parameter and a vorticity parameter as is done, for example, in the 3+1 covariant formulation of Einstein field equations \cite{Elhers, Ellis, Elst, BobSec}. We emphasise that this presentation is for a co-moving velocity and hence single fluid approximation. What if the two neighbouring observers are not co-moving? Could we formulate projection tensors related to two velocities and what could we learn from this? It clear that time parameters would be different $()^{.}=w^{\mu}\tilde{\nabla}_{\mu}()$ and $()^{'}=v^{\mu}\nabla_{\mu}()$, assuming two different velocity vectors $w^{\mu}(\equiv u^{\mu}_{\vert{X}}$) and $v^{\mu}(\equiv u^{\mu}_{\vert{Y}}$). The two spacial derivatives are covariant (along the surfaces as described by the metrics ${g_{\mu\nu}}_{\vert_{w}}$ and ${g_{\mu\nu}}_{\vert_{v}}$ respectively). One could theorise about possible projection tensors $h_{\mu\nu_{\vert{w}}}=f({g_{\mu\nu}}_{\vert_{w}},w_{\mu},w_{\nu})$, $h_{\mu\nu_{\vert{v}}}=f({g_{\mu\nu}}_{\vert_{v}},v_{\mu},v_{\nu})$,  and the intriguing case $h_{\mu\nu_{\vert{wv}}}=f({g_{\mu\nu}}_{\vert_{w,v}},w_{\mu},v_{\nu})$ . The projectors $h_{\mu\nu_{\vert{w}}}$ and $h_{\mu\nu_{\vert{v}}}$ are the familiar tensors found in literature. $h_{\mu\nu_{\vert{wv}}}$ is new and demands further investigation. The fundamental problem in this case is how the two velocities couple to give rise to an effective velocity. There are many different possible configuration that could yield such effective velocity, for example the configuration of non-interacting fluids considered in \cite{Bayin1} where anisotropy was studied. In this article we will present alternative configuration, as an illustration.
 
Let $w_{\mu}$ and $v_{\mu}$ be two 4-velocity unit vectors ( $w^{\mu}w_{\mu}=-1=v^{\mu}v_{\mu}$) that give rise to a complex 4-velocity $u_{\mu}$ defined by
\bea
u_{\mu}&=&w_{\mu}+iv_{\mu}\label{ucomp1}
\eea and whose conjugate is
\bea
u^{\mu}&=&w^{\mu}-iv^{\mu}\label{ucomp2}
\eea where $i=\sqrt{-1}$. This can be found by requiring the existence of Cauchy-Riemann like equations for the 4-dimensional objects, from which a complex potential $\Phi(w,v)$ \cite{BobComplex} may be defined. A complementary scalar potential function was used in formulating the dissipative relativistic fluid theory of the divergence type, the reader is referred to appendix (\ref{div}) for a summary. In our case, we need $u^{\mu}$ to be analytic with respect to a covariant derivative at a given event \cite{Event}. It is then clear from Eqs.({\ref{ucomp1}) and (\ref{ucomp2}) that 
\bea
\label{doubleC}u^{\mu}u_{\mu}=w^{\mu}w_{\mu}+v^{\mu}v_{\mu}=-2
\eea from which we can define the fundamental relation
\bea
\frac{u^{\mu}}{\sqrt{2}}\frac{u_{\mu}}{\sqrt{2}}=-1.
\eea We see in this section that two unit vector in complex configuration generates a unit vector $\hat{u}=u/\sqrt{2}.$ We can now investigate the implication of having such a complex 4-velocity vector.
\subsection{Effective velocity for more fluids species} An important question to address before we continue is how does one construct the effective 4-velocity given more velocity fields. For example, let $w,v,\xi$ and $\zeta$ be 4-velocities in a complex configuration. Three of these give rise to a effective 4-velocity as follows $$u_{\mu}=w_{\mu}+i (v_{\mu}+i\xi_{\mu})=(w_{\mu}-\xi_{\mu})+i v_{\mu}$$ The case of four components is given by
$$u_{\mu}=w_{\mu}+i [v_{\mu}+i(\xi_{\mu}+i\zeta)]=(w_{\mu}-\xi_{\mu})+i (v_{\mu}+\zeta)$$ The extension to a greater number of velocity fields in this formalism should be straight forward. We will restrict our discussion to the case of two fluid species.
\subsection{A projection tensor}\label{proj0}
We begin by constructing a projection tensor onto an emergent surface using the 4-velocity $\hat{u}^{\mu}$ ( where we define $\hat{u}^{\mu} =u^{\mu}/\sqrt{2}$. Such a projection tensor will take the generic form:
\bea
{{\hat{h}}^{\mu}}_{~\nu}={\hat{g}^{\mu}}_{~\nu}+\hat{u}^{\mu}\hat{u}_{\nu},
\eea and is defined to obey the orthogonality condition ${\hat{h}^{\mu}}_{~\nu}\hat{u}^{\mu}=0.$ 
The associated projected tensor ${h^{\mu}}_{\nu}={g^{\mu}}_{\nu}+w^{\mu}w_{\nu},$ projects onto the rest frame only if there is no vorticity. For example, the projection tensor used in {\it Friedmann-Walker} models with perfect fluids matter characterised by local isotropy\cite{Henk}. But the tensor ${\hat{h}^{\mu}}_{~\nu}$ is not the same as ${{h}^{\mu}}_{\nu}$ and does not project onto the hyper-surfaces defined by either $w_{\mu}$ or $v_{\mu}$ but rather to one defined by $\hat{u}^{\mu}$. It is easy to show that that if $w_{\mu}\equiv  v_{\mu}$ then
\bea
u^{\nu}u_{\nu}=v^{\nu}v_{\nu}(1+i)(1-i)=2v^{\nu}v_{\nu}=-2, 
\eea recovering the result in Eq.(\ref{doubleC}). We now have a projection tensor $\hat{h}=\hat{h}({g_{\mu\nu}}_{\vert_{\mu\nu}}, \hat{w}_{\mu},\hat{v}_{\nu})$ which can be resolved into the fundamental velocities; $ w_{\mu}$ and $v_{\nu}$. 

\bea\label{proj}
\hat{h}^{\mu\nu}&=&g^{\mu\nu}+\hat{u}^{\mu}\hat{u}^{\nu}=g^{\mu\nu}+\frac{1}{2}u^{\mu}u^{\nu}\nonumber\\
&=&g^{\mu\nu}+\frac{1}{2}[(w^{\mu}w^{\nu}-v^{\mu}v^{\nu})+i(v^{\mu}w^{\nu}+v^{\nu}w^{\mu})]\nonumber\\
&=&\frac{1}{2}[g^{\mu\nu}+(w^{\mu}w^{\nu}-v^{\mu}v^{\nu})+g^{\mu\nu}+i(v^{\mu}w^{\nu}+v^{\nu}w^{\mu})]\nonumber\\
&=&\frac{1}{2}[g^{\mu\nu}+(w^{\mu}w^{\nu}-v^{\mu}v^{\nu})+ i(-ig^{\mu\nu}+v^{\mu}w^{\nu}+v^{\nu}w^{\mu})]\nonumber\\
&=&\hat{h}_{\mu\nu(R)}+i\hat{h}_{\mu\nu(C)}
\eea where $R$ denotes the real part, while $C$ the complex part.

The presentation above looks deceptively familiar but the project tensor separate velocities not same velocity pointing in different direction as is standard in literature. This distinction should always be kept. We can now ask the question, how does this affect the energy momentum tensor in the case of the general relativistic multi-fluid? We will restrict our discussions to the case of two relativistic fluids. Here we employ non-standard decomposition as will be explained.
\begin{figure}[ht]
\begin{center}
\includegraphics[width=0.35\textwidth]{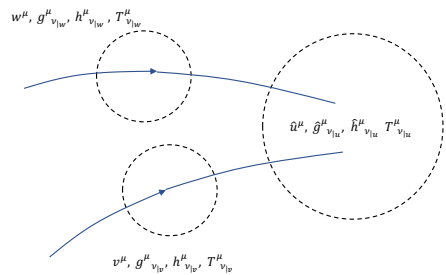} 
\caption{{\it Two original separate velocity vectors, projection tensors and energy momentum tensors and the resultant.\label{fig2}}}
\end{center}
\end{figure}

The substructure of the projection tensor induced by the different 4-velocities (see \ref{proj}) affects the fundamental equations for the multi-fluid system, which in turn affects the momentum, energy and mass equations. In order to grasp the consequences of such sub-structuring, first consider heat conduction. Rather than following the standard bottom-up approach where the equation representing the bulk behaviour is derived from sub-constituents states, we adopt a top-down approach. In this case we decompose the equations for the bulk into the constituent parts.Let's begin with the heat vector whose divergence contributes to the energy balance equation.

\subsection{ The heat flux vector}\label{heateq}
Using the projector tensor $\hat{h}_{\mu\nu}$, we can define a heat vector with respect to the velocity $\hat{u}_{\gamma}.$ We find,
\bea
\hat{q}_{\mu}&=&-\hat{h}_{\mu\nu}\hat{u}_{\gamma}T^{\nu\gamma}\equiv-\hat{h}_{\mu\nu}\left(\frac{u_{\gamma}}{\sqrt{2}}\right)T^{\nu\gamma},\nonumber\\
&=&-\frac{1}{\sqrt{2}}(\hat{h}_{\mu\nu(R)}+i\hat{h}_{\mu\nu(C)})\left(w_{\gamma}+i v_{\gamma}\right)T^{\nu\gamma},\nonumber\\
&=&-\frac{1}{\sqrt{2}}\left[(\hat{h}_{\mu\nu(R)}w_{\gamma}-\hat{h}_{\mu\nu(C)}v_{\gamma})\right]T^{\nu\gamma}\nonumber\\&&~~~~~~~~~~~~~~~
-\frac{1}{\sqrt{2}}i\left[(\hat{h}_{\mu\nu(C)}w_{\gamma}+\hat{h}_{\mu\nu(R)}v_{\gamma})\right]T^{\nu\gamma},\nonumber\\
\eea 

This shows the heat vector is composed of a real and a complex part. It follows that heat vector's contribution to the stress momentum tensor is given by
\bea\label{q1}
\hat{q}_{(\mu}\hat{u}_{\nu)}&=&-\hat{h}_{(\mu\gamma}\hat{u}_{\delta}T^{\gamma\delta}\hat{u}_{\nu)}=-\frac{1}{\sqrt{2}}(\hat{h}_{(\mu\nu}\hat{u}_{\delta})T^{\nu\delta}(w_{\nu)}+i v_{\nu)}),\nonumber\\
\eea where each term with a hat, $\hat{f}$, can be decomposed into real and complex parts respectively. Eqs. (\ref{q1}) and (\ref{qu1}), together yield

\bea\label{qu1}
\hat{q}_{(\mu}\hat{u}_{\nu)}&=&\hat{h}_{(\mu\gamma}\hat{u}_{\delta}T^{\gamma\delta}\hat{u}_{\nu)}\nonumber\\
&=&\frac{1}{2}\left[(\hat{h}_{\mu\nu(R)}w_{\delta}w_{\nu)}-\hat{h}_{\mu\nu(C)}v_{\delta}w_{\nu)})\right]T^{\nu\delta}\nonumber\\&&~~
-\frac{1}{2}\left[(\hat{h}_{\mu\nu(C)}w_{\delta}v_{\nu)}+\hat{h}_{\mu\nu(R)}v_{\delta}v_{\nu)})\right]T^{\nu\delta}\nonumber\\&&~~~~~~~
+\frac{1}{2}i\left[(\hat{h}_{\mu\nu(R)}w_{\delta}v_{\nu}-\hat{h}_{\mu\nu(C)}w_{\nu}v_{\delta})\right]T^{\nu\delta}\nonumber\\&&~~~~~~~~
+\frac{1}{2}i\left[(\hat{h}_{\mu\nu(C)}w_{\delta}w_{\nu)}+\hat{h}_{\mu\nu(R)}v_{\delta}w_{\nu)})\right]T^{\nu\delta},\nonumber\\
\eea

\subsection{Energy Density and Pressure Terms}\label{dens} 
The energy density is constituted as follows
\bea
\label{densityC}\label{densityC1}
\hat{\rho}&=&\hat{u}_{\mu}\hat{u}_{\nu}T^{\mu\nu}
\equiv\frac{1}{2}(w_{\mu}w_{\nu}-v_{\mu}v_{\nu})T^{\mu\nu}+\frac{i}{2}( w_{\mu}v_{\nu}+w_{\mu}v_{\nu})T^{\mu\nu}\nonumber\\
\eea Likewise the isotropic pressure term is given by 
\bea
\hat{p}&=&\frac{1}{3}\hat{h}_{\mu\nu}T^{\mu\nu}=\frac{1}{3}(\hat{h}_{\mu\nu(R)}+i\hat{h}_{\mu\nu(C)})T^{\mu\nu}.
\eea The anisotropic pressure is given by is 

\bea
\hat{\pi}_{\mu\nu}&=&T^{\gamma\delta}(\hat{h}_{\gamma\langle \mu(R)}\hat{h}_{\nu\rangle \delta(R)}-\hat{h}_{\gamma\langle \mu(C)}\hat{h}_{\nu\rangle \delta(C)})\nonumber\\~~~~~~~~~~~~&&
+iT^{\gamma\delta}[(\hat{h}_{\gamma\langle\mu(R)}\hat{h}_{\nu\rangle\delta(C)})+(\hat{h}_{\gamma\langle\mu(C)}\hat{h}_{\nu\rangle\delta(R)})],\nonumber\\
\eea
where the real and complex parts are clearly manifest. 

\subsection{\label{EMT}The Energy Momentum Tensor}
The total stress-energy momentum tensor takes the form
\bea
T_{\mu\nu}&=&\hat{\rho}\hat{u}_{\mu}\hat{u}_{\nu}+\hat{p}\hat{h}_{\mu\nu}+2\hat{u}_{(\mu}\hat{q}_{\nu)}+\hat{\pi}_{\mu\nu},
\eea which structurally resembles the standard single-fluid form but hides the real and complex constituents. An observer living in the real plane, with no knowledge of the existence of the complex plane, will measure a total energy density that does not match what they expect i.e. $(w_{a}w_{b}-v_{a}v_{b})T^{ab}/2$ as seen in Eq. (\ref{densityC1}) instead of $w_{a}w_{b}T^{ab}$ given by the own 4-velocity $w^{a}$. This disparity between the expected and the observed measures may help account for some of the disparities between predictions from {\it single-fluid approximation} and what is observed in cosmology. In general, this form of flow is anisotropic and may provide test-ground for the cosmological principle. In our case, we have considered the limit in which the system isotropizes.
 
The illustrative velocity of $\hat{u}_{c}$ is but one example of how two fluids could be coupled. It belongs to a family of couplings that are expressed as  $\hat{u}_{C}=f(w_{C}, v_{C},...)$ where $'..'$ expresses the fact that there may be more velocity fields. We know that other configurations \cite{Bayin1, Bayin2, Letelier} have been used to study anisotropic models where the energy-stress tensor is primarily mattered tensors. In this article, we consider such couplings in the context of thermodynamics.

\subsection{The Entropy vector}
There are two formal definitions of entropy: the thermodynamics and the statistical. We take the thermodynamic viewpoint. In this regards, the classical thermodynamics theory considers a system that is composed of constituents, but whose state is found by taking the averages of thermodynamic properties of such constituent; in effect looking at the cumulative behaviour. Although the initial development of the concept only considered such averages for a system that was in equilibrium via statistical mechanics, the latter development extends the theory by incorporating aspects that allowed for the non-equilibrium thermodynamics via the kinetic theory. It is the latter version that is of interest to us. 
Standard treatment of statistical thermodynamics, see \cite{Callen}, is based on postulates that are given in terms of the behaviour of simple systems. These are systems microscopically homogeneous, isotropic and devoid of electric charge, chemical reactions, electrical force fields or surface effects. In order to account for multi-fluids, where some of these properties cannot be neglect, it is imperative that we go beyond the simple system postulates. To this end, we follow \cite{Isra0, Isra1, Muller1} who, by incorporate the quadratic terms in the heat flux and viscous stresses in the expression for the entropy 4-vector, obtain a generalised theory able to describe transient non-equilibrium thermodynamics satisfying the causality condition.
It is straightforward to show that the entropy current takes for this flow takes the form:

\bea\label{en1}
S^{\mu}_{\vert i}&=&\hat{s}\hat{u}^{\mu}_{\vert i}+\hat{s}{\mu}_{\vert i}\nonumber\\&\equiv&\hat{s}\hat{u}^{\mu}_{\vert i}+{\hat{q^{\mu}_{\vert i}}}\frac{1}{T}-\left(\hat{\beta}_{0}\Pi^2+\hat{\beta}_{1}\hat{q}_{\nu\vert i}\hat{q}^{\nu}_{\vert i}+\hat{\beta}_{2}\hat{\pi}_{\gamma\delta\vert i}\hat{\pi}^{\gamma\delta\vert i}\right)\frac{\hat{u}^{\mu}_{\vert i}}{2T}\nonumber\\&&~~~~~~~~~~~~~~~
+\left(\hat{\alpha}_{0}\Pi \hat{q}^{\mu}_{\vert i}+\hat{\alpha}_{1}\hat{\pi}_{\mu\nu\vert i}\hat{q}^{\nu}_{\vert i}\right)\frac{1}{T}.\nonumber\\
\eea
here too, any term with a hat, $\hat{f}$, can be expanded in terms of a real  part and an imaginary part. $\hat{s}$ is the entropy density, $\hat{s}^{\mu}$ is the entropy flux with respect to $\hat{u}^{\mu\vert i}$ such that $\hat{s}_{\mu\vert i}\hat{u}^{\mu}_{\vert i}=0$. $\Pi$ is the bulk viscosity. Here we have considered the limit in which the temperature is universally \cite{BobTim, BobTemp}. The coefficients $\hat{\beta}_{0}, \hat{\beta}_{1}, \hat{\beta}_{2}, \hat{\alpha}_{0}$ and $\hat{\alpha}_{1}$ are the generalized case  of the counterparts appearing in the MIS theory.

\section{ Discussion and Conclusion}\label{sec4}
We have developed a generic expression for stress-energy-momentum tensor and entropy taking into account a multi-fluid configuration. The formulation extends the $MIS$ theory by incorporating more than one fluid species. Our starting point is the construction of the {\it effective} 4-velocity $\hat{u}^{\mu}$ that is the resultant of the various fluid species velocities and which is defined by the Cauchy-Riemann equations for 4-dimensions \cite{BobComplex}. The standard approach in modelling non-conducting fluids species uses the velocity of the centre of mass or gravity as the representative velocity and the whole fluid is then treated as a single fluid. In contrast, we use a complex formulation which allows us to retain and to monitor the unique or peculiar contributions from individual species. This allows for the treatment of bulk behaviour \cite{Pad3} and that incorporates fluid interactions that may or may not as expressed by Eq. (\ref{Emt1}). This is particularly important for the treatment of well-behaved heat conduction in relativistic fluids that includes dissipation \cite{Carter1, Andersson1}

So what is the utility of this formulation? It is thought that the recently discovered late time acceleration of the universe could be explained by invoking  $dark~energy$ whose density is usually added into the Friedman equation by hand, without a hint of its source. We think this can be remedied. In the illustration, we have considered a formulation that brings two fluids together whose effective 4-velocity can be expressed as a complex vector. This plays two roles, (i) it ensures that the fluids remain distinct and (ii) it allows for two fluids to have an impact on the other. From an observer point of view, the one living in the $real$ part will not have any knowledge of the existence of the complex dynamics but will notice that the total energy density is not what they expect i.e. $\rho_{\vert_{X}}=w_{\mu}w_{\nu}T^{\mu\nu}$, but one that is modulated by some mysterious addition $(w_{\mu}w_{\nu}-v_{\mu}v_{\nu})T^{\mu\nu}/2\equiv\rho_{\vert_{X}}+\rho_{\vert_{Y}}$. The negative sign here is an artefact of the complex analysis not necessarily bearing direct physical significance. The observer in the imaginary hyper-surface will, similarly, notice the difference in their energy density. This hints at a $twin-universe$ theory and by extension a multi-verse theory. Since there are numerous ways of formulating an effective 4-velocity, it is clear that our formulation belongs to a family of such and demands further investigation.
\section*{Acknowledgement}
The author acknowledges funding support from the University of Cape Town's Next professoriate Programme (NGP). 
\appendix
\section{A generic approach: The dissipative relativistic fluid theories of divergence type}\label{div}
Let the label $X$ denote a relativistic dissipative fluid moving with a 4-velocity $w^{c}$\cite{Andersson1, Rezzolla1}. We have decided to keep the index $X$ as a reminder that we have a single fluid that is identified by this label. Let the fluid satisfy the following properties:
\begin{enumerate}
\label{prop}\item[a)] The dynamical variables are the individual particle-number currents vectors, $N^{\mu}_{X}$ and the total stress-energy momentum tensor $T^{\mu\nu}_{X}$. %
\item[b)] The conservation laws in addition to the dynamical equations are:
\bea
\nabla_{\mu}N^{\mu}_{X}&=&0\\
\nabla_{\mu}T^{\nu\mu}_{X}&=&0,\\
\nabla_{\mu}A^{\mu\nu\delta}_{X}-I^{\nu\delta}_{X}&=&0,
\eea
where the covariant derivative based on a space-time geometry endow with the metric tensor $g^{\mu\nu}$. It is known \cite{Notes} that covariant derivatives on space-time and the covariant derivatives defined in the geometry of a space-like hyper-surface in space-time are linked through the projection of a space-like connection. $A$ and $I$ are the algebraic forms of symmetric and the trace-free  $N^{\nu}_{X}$ and $T^{\nu\mu}$ such that 
\item[c)] There exists a combined entropy current $s^{\nu}_{X}$ that satisfies 
\bea
\nabla_{\nu}s^{\nu}_{X}=\sigma,
\eea
where $\sigma$ is some algebraic form of $N^{\nu}_{X}$ and $T^{\nu\mu}_{{X}}$. Reversible or thermodynamic equilibrium state is given when $\sigma=0$, while the irreversible state is when $\sigma>0.$ 
\end{enumerate}
Theories that satisfy these condition as called divergence theories. 
\bea
h^{\mu\nu} _{\vert{X}}={g^{\mu\nu}_{\vert{X}}}+u^{\mu}_{\vert{X}}u^{\nu}_{\vert{X}}.
\eea It is important to note that the resultant velocity $u^{\nu}_{\vert{X}}$ is a generic vector function whose form is dependent on the  velocities of the two fluid species; an example was considered in the previous section. 

The reader will note that the presentation above extends those of \cite{Geroch1} to include two coupled particle types. Indeed, this can be generalised to include more particle species. In the two-species case, the general theory obeying the three properties above may be shown to be generated by the scalar potential ; $\chi_{X}$ and a tensor $I^{\nu\mu}_{X}$ 
\bea
N^{\mu}_{X}&=&\frac{\partial^2\chi_{X}}{\partial\xi\partial\xi_{\mu}}\\
T^{\mu\nu}_{X}&=&\frac{\partial^2\chi_{X}}{\partial\xi_{\mu}\partial\xi_{\nu}}\\
A^{\nu\mu\gamma}_{X}&=&\frac{\partial^2\chi_{X}}{\partial\xi_{\mu}\partial\xi_{\nu\gamma}}
\eea with the entropy current given by
\bea
\label{first}S^{\nu}&=&\left(\frac{\partial\chi_{X}}{\partial\xi_{\nu}}-\xi N^{\nu}_{X}\right)-\xi_{\mu}T^{\nu\mu}_{X}-\xi_{\mu\gamma}A^{\nu\mu\gamma}_{X},\\
\label{second}\sigma&=&-\xi_{\nu\mu}I^{\nu\mu}. 
\eea Equation (\ref{second}) is the result of taking the divergence of equation (\ref{first}) given the three properties (\ref{prop}). It is straight forward to show that this is the general theory satisfying these properties. The velocity fields are chosen in such a way that the entropy density satisfy the concavity requirement \cite{Muller1}. The selection is non trivial in general relativity since the definition of entropy density is frame dependent; which in itself is not unique. In order to make progress, we assume that privileged frames, satisfying the above requirements, exists \cite{Rug}. In this regard, the stress-energy momentum tensor may be written in the form
\bea
T^{\mu\nu}_{X}&=&\left(\rho_{X}u^{\mu}u^{\nu}+p_{X}h^{\mu\nu}+2u_{X}^{(\mu}q_{X}^{\nu)}+\pi^{\mu\nu}_{X}\right),
\eea
\bea
N^{\mu}_{X}&=&n_{X}u_{X}^{\mu}+n_{X}^{\mu},
\eea such that $n^{\nu}_{X}u_{X\nu}=0.$
The description is  generic, but can easily be adapted to recover some of the known theories. For example, Eckart's theory is easily recovered when one restricts the above formulation to first order approximation and the $MIS$ theory in the previous section for second order approximation  (see \cite{Geroch1}). 
These formulations are useful for modelling causally well-behaved heat conduction in relativistic fluids \cite{Carter1, Andersson1} in the context of single fluid approximation. 
The approach above is based on defining a scalar-type generating function that gives rise to the fundamental vector and tensor dynamical variables. These parameters are locally defined and allow for the notion of a {\it thermodynamics-equilibrium}. The technique is not dissimilar to the decomposition of cosmological perturbations into local scalar-type, local vector-type and local tensor-type \cite{Clarkson1}. This in contrast to the standard Helmholtz's theorem used to define non-local scalars and vectors \cite{Stewart1, Kodama1, Stewart2}. The challenge with non-local variables is that one needs to specify boundary conditions in order to define such variables. Although it is often difficult to map theories derived from locally defined variables to those not based on non-local variables, it possible to find a set that allows for such links. For example, it is possible to reconstruct a non-local theory, using divergence derivatives of a scalar variable, into a corresponding local theory as demonstrated in \cite{Clarkson1}. In this regard the scalar variable, or better still a function, becomes a generator of the specified field theory.

\end{document}